\documentclass[12pt]{article}
\usepackage{amssymb,amsmath,epsfig}

\begin{document}
\title{\bf Dynamics of Bianchi
$I$ Universe with Magnetized Anisotropic Dark Energy}

\author{M. Sharif \thanks{msharif@math.pu.edu.pk} and M. Zubair
\thanks{mzubairkk@gmail.com}\\\\
Department of Mathematics, University of the Punjab,\\
Quaid-e-Azam Campus, Lahore-54590, Pakistan.}

\date{}

\maketitle
\begin{abstract}
We study Bianchi type $I$ cosmological model in the presence of
magnetized anisotropic dark energy. The energy-momentum tensor
consists of anisotropic fluid with anisotropic EoS $p=\omega{\rho}$
and a uniform magnetic field of energy density $\rho_B$. We obtain
exact solutions to the field equations using the condition that
expansion is proportional to the shear scalar. The physical behavior
of the model is discussed with and without magnetic field. We
conclude that universe model as well as anisotropic fluid do not
approach isotropy through the evolution of the universe.
\end{abstract}


\section{Introduction}

Recent cosmological observations contradict the matter dominated
universe with decelerating expansion indicating that our universe
experiences accelerated expansion. The accelerating expansion of the
universe is driven by mysterious energy with negative pressure known
as \emph{Dark Energy} (DE). The evidence of the existence of DE
comes from the Supernova observations \cite{1,2} and other
observations such as cosmic microwave background (CMB) anisotropies
measured with WMAP satellite \cite{3} and large scale structure
\cite{4}. These observations suggest that nearly two-third of our
universe consists of DE and the remaining consists of relativistic
dark matter and baryons \cite{5}.

In spite of all the observational evidences, the nature of DE is
still a challenging problem in theoretical physics. A variety of
possible solutions such as cosmological constant \cite{6},
quintessence \cite{7}, phantom field \cite{8}, tachyon field
\cite{9}, quintom \cite{10}, and the interacting DE models like
Chaplygin gas \cite{11}, holographic models \cite{12} and braneworld
models \cite{13} etc. have been proposed to interpret accelerating
universe. However, none of these models can be regarded as being
entirely convincing so far.

Recently, many authors have studied the Bianchi type $I$ model in
the presence of anisotropic DE. Rodrigues \cite{14} constructed a
Bianchi type $I$ ${\Lambda}$CDM cosmological model whose DE
component preserves non-dynamical character but yields anisotropic
vacuum pressure. Koivisto and Mota \cite{15,16} proposed a different
approach to resolve the CMB anisotropy problem; the earlier isotropy
of the universe could be distorted by the direction dependent
acceleration of the later universe. Koivisto and Mota \cite{16}
investigated the Bianchi $I$ cosmological model containing
interacting DE fluid with non-dynamical anisotropic EoS and perfect
fluid component. They suggested that if the EoS is anisotropic, the
expansion rate of the universe becomes direction dependent at late
times and cosmological models with anisotropic EoS can explain some
of the observed anomalies in CMB.

Mota et al. \cite{17} explored the possibility of using the
cosmological observation to probe and constrain an imperfect DE
fluid. They concluded that a perfect fluid representation of DE
might ultimately turn out to be a phenomenologically sufficient
description of all the observational consequences of DE. However,
one cannot exclude the possibility of imperfectness in DE. Akarsu
and Kilinc \cite{18,19} suggested that anisotropic fluid must not
necessarily promote anisotropy in the expansion whereas such fluid
may also act to support isotropic behavior of the universe. It has
been shown \cite{18} that anisotropic Bianchi $I$ model in the
presence of perfect fluid and minimally interacting DE shows
isotropic behavior for the earlier times of the universe.

Primordial magnetic fields can have a significant impact on the CMB
anisotropy depending on the direction of field lines \cite{20,21}.
Many people have investigated the influence of magnetic field on the
dynamics of universe by analyzing anisotropic Bianchi models.
Milaneschi and Fabbri \cite{22} studied the anisotropy and
polarization properties of CMB radiation in homogeneous Bianchi $I$
cosmological model. Jacobs \cite{23} explored the effects of a
uniform, primordial magnetic field on Bianchi type $I$ cosmological
model. He concluded that the primordial magnetic field produced
large expansion anisotropies during the radiation-dominated phase
but it had negligible effect during the dust-dominated phase. King
and Coles \cite{21} discussed the dynamics of magnetized
axisymmetric Bianchi $I$ universe with vacuum energy. He examined
the behavior of scale factors perpendicular and parallel to the
field lines. Roy et al. \cite{24} investigated Bianchi type $I$
cosmological models containing perfect fluid and magnetic field
directed along $x$ axis. Exact solutions are obtained using the
condition that expansion is proportional to shear scalar.

In this paper, we would like to investigate the effects of magnetic
field on the dynamics of anisotropic Bianchi $I$ model in the
presence of anisotropic DE. The paper has the following format. In
section \textbf{2}, we present anisotropic Bianchi type $I$ model
and formulate the dynamical field equations which describe the
evolution of the universe. In section \textbf{3}, we obtain exact
solution to the field equations and discuss the physical properties
of the solution. Section \textbf{4} contains a brief discussion
related to two special cases for $\beta=0$ and $m=1$. Finally, in
section \textbf{5}, we summarize the results.

\section{Bianchi $I$ Model and the Field Equations}

The line element for the spatially homogeneous, anisotropic and
LRS Bianchi type $I$ spacetime is given by
\begin{equation}\label{1}
ds^{2}=dt^2-A^2(t)dx^2-B^2(t)(dy^2+dz^2),
\end{equation}
where the scale factors $A$ and $B$ are functions of cosmic time
$t$ only. For $A(t)=B(t)=a(t)$, this reduces to the flat
FRW spacetime. This spacetime has one
transverse direction $x$ and two equivalent longitudinal
directions $y$ and $z$. We assume that the universe is filled with
anisotropic fluid, and that there is no electric field while the
magnetic field is oriented along $z$ axis. The scale factor
$A(t)$ is in the transverse direction, perpendicular to magnetic
field while $B(t)$ is along the direction of field lines. King and
Coles \cite{21} and Jacobs \cite{23} used the magnetized perfect
fluid energy-momentum tensor to discuss the effects of magnetic
field on the evolution of the universe.

Here we take a more general energy-momentum tensor for the
magnetized anisotropic DE fluid in the following form
\begin{equation}\label{2}
T_{\mu}^{\nu}=diag[\rho+{\rho}_B,-p_{x}+{\rho}_B,-p_{y}-{\rho}_B,-p_{z}-{\rho}_B],
\end{equation}
where $\rho$ is the energy density of the fluid; $p_x,~p_y$ and
$p_z$ are pressures on $x,~y$ and $z$ axes respectively and
$\rho_B$ stands for energy density of magnetic field. The
anisotropic fluid is characterized by the EoS $p=\omega{\rho}$,
where $\omega$ is not necessarily constant \cite{25}. From
Eq.(\ref{2}), we have
\begin{equation}\label{3}
T_{\mu}^{\nu}=diag[\rho+{\rho}_B,-({\omega}+{\delta})\rho+{\rho}_B,-({\omega}+{\gamma})
\rho-{\rho}_B,-({\omega}+{\gamma})\rho-{\rho}_B],
\end{equation}
where ${\omega}_x={\omega}+{\delta},~{\omega}_y={\omega}+{\gamma}$,
and ~${\omega}_{z}={\omega}+{\gamma}$ are the directional EoS
parameters on $x,~y$ and $z$ axes respectively. $\delta$ and
$\gamma$ are the deviations from $\omega$ on $x$ and $y$, $z$ axes
respectively. If the deviation parameters are zero, then
Eq.(\ref{2}) represents the energy-momentum tensor for the isotropic
fluid and magnetic field \cite{21}. For zero magnetic field,
Eq.(\ref{2}) is reduced to the energy-momentum tensor of anisotropic
fluid \cite{18}.

The Einstein field equations are given by
\begin{equation}\label{4}
G_{{\mu}{\nu}}=R_{{\mu}{\nu}}-\frac{1}{2}Rg_{{\mu}{\nu}}
=T_{{\mu}{\nu}},
\end{equation}
where $R_{{\mu}{\nu}}$ is the Ricci tensor, R is the Ricci scalar
and $T_{{\mu}{\nu}}$ is the energy-momentum tensor for magnetized
anisotropic fluid. For the Bianchi type $I$ spacetime, the field equation take the form
\begin{eqnarray}\label{5}
2\frac{\dot{A}\dot{B}}{AB}+\frac{\dot{B}^2}{B^2}&=&{\rho}+{\rho}_B,
\\\label{6} 2\frac{\ddot{B}}{B}+\frac{\dot{B}^2}{B^2}&=&-({\omega}+{\delta}){\rho}
+{\rho}_B, \\\label{7}\frac{\ddot{A}}{A}+
\frac{\ddot{B}}{B}+\frac{\dot{A}\dot{B}}{AB}&=&-({\omega}+{\gamma}){\rho}-{\rho}_B,
\end{eqnarray}
where dot denotes derivative with respect to time. The energy
conservation equation, $T^{\mu}_{\nu;\mu}=0$,
leads to two equations for the anisotropic fluid and magnetic
field \cite{21}
\begin{eqnarray}\label{9}
\dot{\rho}+(1+\omega)\rho(\frac{\dot{A}}{A}+2\frac{\dot{B}}{B})+
\rho(\delta\frac{\dot{A}}{A}+2\gamma\frac{\dot{B}}{B})&=&0,\\\label{10}
{\rho}_B&=&\frac{\beta}{B^4}.
\end{eqnarray}

The conservation equation for the anisotropic fluid can be
decomposed into two parts,
\begin{equation}\label{11}
T^{\mu}_{\nu;\mu}=\acute{T}^{\mu}_{\nu;\mu}+{\tau}^{\mu}_{\nu;\mu}=0,
\end{equation}
where ${\tau}^{\mu}_{\nu;\mu}$ is the last term in Eq.(\ref{9})
which arises due to the anisotropy in the fluid and
$\acute{T}^{\mu}_{\nu;\mu}$ represents the deviation free part of
the $T^{\mu}_{\nu;\mu}$. Let us take \cite{18}
\begin{equation}\label{12}
{\tau}^{\mu}_{\nu;\mu}=\rho(\delta\frac{\dot{A}}{A}+2\gamma\frac{\dot{B}}{B})=0.
\end{equation}
Using this assumption in Eq.(\ref{9}), we obtain conservation of the
perfect fluid
\begin{equation}\label{13}
\acute{T}^{\mu}_{\nu;\mu}=\dot{\rho}+(1+\omega)\rho(\frac{\dot{A}}{A}+2\frac{\dot{B}}{B})=0.
\end{equation}
Equation (\ref{12}) is satisfied either $\delta(t)$ and $\gamma(t)$
are trivially zero or the ratio of expansion rate on $x$ axis to the
$y$ axis is equal to ${-2\gamma}/{\delta}$. In order to obtain more
general solution, the deviation parameter on the $x$ axis
$\delta(t)$ is assumed to be \cite{18}
\begin{equation}\label{14}
\delta(t)=n\frac{2}{3}\frac{\dot{B}}{B}(\frac{\dot{A}}{A}+2\frac{\dot{B}}{B})\frac{1}{\rho}
\end{equation}
and hence the deviation parameter on $y$ and $z$ axes is given by
\begin{equation}\label{15}
\gamma(t)=-n\frac{1}{3}\frac{\dot{A}}{A}(\frac{\dot{A}}{A}+2\frac{\dot{B}}{B})\frac{1}{\rho},
\end{equation}
where $\delta(t)$ and $\gamma(t)$ are dimensionless parameters and
$n$ is the real dimensionless constant that parameterizes the
deviation from EoS parameter. The anisotropy of the DE is measured
using the relation $(\delta(t)-\gamma(t))/\omega(t)$ and for $n=0$,
DE is found to be isotropic.

\section{General Parameters and Solution of the Field Equations}

Here we define some parameters for the Bianchi $I$ model which are
important in cosmological observations. The average scale factor
and the volume are defined as
\begin{equation}\label{16}
a=(AB^2)^{\frac{1}{3}},\quad V=a^3=AB^2.
\end{equation}
The anisotropy parameter of the expansion is characterized by the
mean and directional Hubble parameters and is defined as
\begin{equation}\label{17}
\Delta=\frac{1}{3}\sum_{i=1}^3(\frac{H_i-H}{H})^2,
\end{equation}
where
\begin{equation*}
H=\frac{1}{3}({\ln}V\dot{)}={\ln}\dot{a}
=\frac{1}{3}(\frac{\dot{A}}{A}+2\frac{\dot{B}}{B}),
\end{equation*}
is the mean Hubble parameter and $H_i(i=1,2,3)$ represent the
directional Hubble parameters in the directions of $x,~y$ and $z$
axes respectively, and are given by
\begin{equation*}
H_x=\frac{\dot{A}}{A}, \quad H_y=H_z=\frac{\dot{B}}{B}.
\end{equation*}
The anisotropy of the expansion results in isotropic expansion of
the universe for $\Delta=0$. The physical parameters like scalar
expansion $\Theta$, shear scalar $\sigma^2$ are given by
\begin{eqnarray}\label{18}
\Theta&=&u_{; a}^a=\frac{\dot{A}}{A}+2\frac{\dot{B}}{B},
\\\label{19} \sigma^2&=&\frac{1}{2}\sigma_{ab}\sigma^{ab}=
\frac{1}{3}[\frac{\dot{A}}{A}-\frac{\dot{B}}{B}]^2.
\end{eqnarray}
It is mentioned here that any universe model becomes isotropic for
the diagonal energy-momentum tensor when
$t\rightarrow{+\infty},~\Delta\rightarrow0,~
V\rightarrow{+\infty}$ and $T^{00}>0~(\rho>0)$ \cite{19,26}.

In order to solve the field equations, we use a physical condition
that expansion scalar is proportional to shear scalar. According to
Throne \cite{27}, observations of velocity red shift relation for
extragalactic sources suggest that Hubble expansion of the universe
is isotropic within about 30\% range approximately \cite{28,29} and
red shift studies place the limit $\frac{\sigma}{H}{\leq}.30$, where
$\sigma$ is the shear and $H$ is the Hubble constant. Collins
\cite{30} discussed the physical significance of this condition for
perfect fluid and barotropic EoS in a more general case. In many
papers \cite{24,31}-\cite{33}, this condition is proposed to find
the exact solutions of cosmological models. It is given by
\begin{equation}\label{20}
A=B^m,
\end{equation}
where $m\neq1$ is a positive constant. Subtracting Eqs.(\ref{6})
and (\ref{7}), it follows that
\begin{equation}\label{21}
\frac{\ddot{A}}{A}+\frac{\dot{A}\dot{B}}{AB}-\frac{\ddot{B}}{B}
-\frac{\dot{B}^2}{B^2}
=\frac{n}{3}(\frac{\dot{A}^2}{A^2}+4\frac{\dot{A}\dot{B}}{AB}
+\frac{\dot{B}^2}{B^2})-\frac{2\beta}{B^4}.
\end{equation}
Applying the condition given in Eq.(\ref{20}), we have
\begin{equation}\label{22}
2\ddot{B}+\frac{2(3m^2-3-nl)}{3(m-1)}\frac{\dot{B}^2}{B}=-\frac{4\beta}{B^3(m-1)},
\end{equation}
where $l=m^2+4m+4$ is a positive constant. Replacing
$\dot{B}=f(B)$, it follows that
\begin{equation}\label{23}
\frac{df^2}{dB}+\frac{2(3m^2-3-nl)}{3(m-1)}\frac{f^2}{B}=-\frac{4\beta}{B^3(m-1)}
\end{equation}
which has the solution
\begin{equation}\label{23a}
f^2=(\frac{dB}{dt})^2
=cB^{-\frac{2(3m^2-3-nl)}{3(m-1)}}-\frac{6{\beta}B^{-2}}{(3m^2-3m-nl)},
\end{equation}
where $c$ is a constant of integration. Thus the spacetime reduces
to the form
\begin{eqnarray}\label{24}
ds^{2}&=&(\frac{dt}{dB})^2{dB}^2-B^{2m}(t)dx^2-B^2(t)(dy^2+dz^2),\nonumber\\
&=&\frac{dT^2}{cT^{-\frac{2(3m^2-3-nl)}{3(m-1)}}-\frac{6{\beta}T^{-2}}{(3m^2-3m-nl)}}
-T^{2m}dx^2-T^2\nonumber\\
&\times&(dy^2+dz^2).
\end{eqnarray}
where $B=T,~x=X,~y=Y,z=Z$.

\subsection{Some Physical Features of the Model}

Now we discuss some physical aspects of the model (\ref{24}). The
directional and the mean Hubble parameters will become
\begin{eqnarray}\label{26}
H_x&=&mH_y=m\{cT^\frac{-2(3m^2+3m-6-nl)}{3(m-1)}-\frac{6{\beta}T^{-4}}{3m^2-3m-nl}\}^{1/2},\nonumber\\
H&=&\frac{m+2}{3}\{cT^\frac{-2(3m^2+3m-6-nl)}{3(m-1)}-\frac{6{\beta}T^{-4}}{3m^2-3m-nl}\}^{1/2}.
\end{eqnarray}
We see that these quantities are found to be dynamical and take
infinitely large values at $T=0$ for $m\geq2$. The values of
Hubble parameters $H,~H_x,~H_y$ decrease with the increase in time
and approach to zero as $T\rightarrow{\infty}$ for
$n<\frac{3m^2+3m-6}{l}$. The scale factors are found to be zero at
$T=0$ and hence the model exhibits \emph{point type} singularity. The
volume of the universe model is given by
\begin{equation}\label{25}
V=T^{m+2}
\end{equation}
and anisotropy parameter of the expansion is found to be
\begin{equation}\label{27}
\Delta=2\frac{(m-1)^2}{(m+2)^2}.
\end{equation}
The expansion and shear scalar take the form
\begin{eqnarray}\label{28}
\Theta&=&3H=(m+2)\{cT^\frac{-2(3m^2+3m-6-nl)}{3(m-1)}
-\frac{6{\beta}T^{-4}}{3m^2-3m-nl}\}^{1/2},\\\label{29}
\sigma&=&\frac{(m-1)^2}{3}\{cT^\frac{-2(3m^2+3m-6-nl)}
{3(m-1)}-\frac{6{\beta}T^{-4}}{3m^2-3m-nl}\}.
\end{eqnarray}

We note that spatial volume is zero at initial epoch and increases
as $T\rightarrow{\infty}$. The expansion and shear scalar are
infinite at $T=0$ and decreases with the increase in cosmic time.
Thus the universe starts evolving with the zero volume at the
initial epoch with infinite rate of expansion which slows down for
the later times of the universe. The component of magnetic field
reduces the expansion, shear scalar and Hubble parameters. The
anisotropy parameter of the expansion is found to be constant and
becomes zero at $m=1$ (it will be discussed as a special case). Thus
the model does not approach to isotropy for the future evolution of
the universe. The most general form of energy density is found by
using Eqs.(\ref{5}) and (\ref{17}) as
\begin{equation}\label{30}
\rho=3H^2(1-\frac{\Delta}{2})-\frac{\beta}{B^4}.
\end{equation}
The energy density of the DE for the model (\ref{24}) turns out to be
\begin{equation}\label{31}
\rho=c(2m+1)T^\frac{-2(3m^2+3m-6-nl)}{3(m-1)}-
\frac{{\beta}T^{-4}(3m^2+9m+6-nl)}{3m^2-3m-nl}.
\end{equation}
The energy condition $\rho{\geq}0$ leads to
\begin{equation*}
T^\frac{-2(3m^2+3m-6-nl)}{3(m-1)}~{\geq}~
\frac{{\beta}T^{-4}(3m^2+9m+6-nl)}{c(2m+1)(3m^2-3m-nl)}.
\end{equation*}
\begin{figure}
\centering \epsfig{file=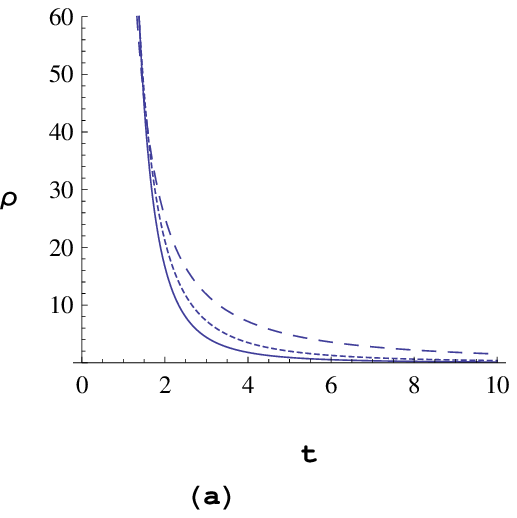,width=.48\linewidth, height=3in}
\epsfig{file=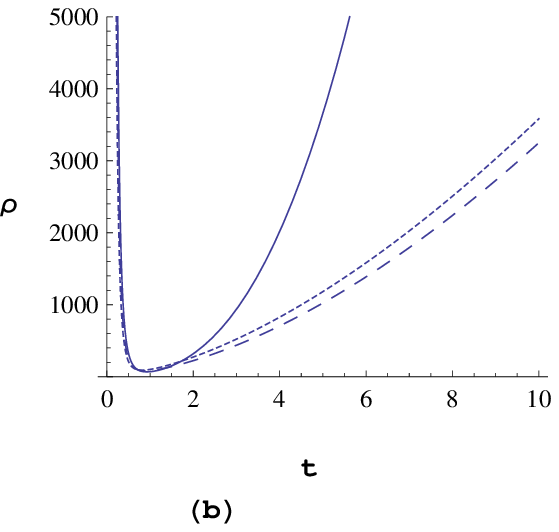,width=.48\linewidth, height=3in}
\caption{Plots of $\rho$ verses cosmic time $t$ (a)
$n<\frac{3m^2+3m-6}{l}$ (b) $n>\frac{3m^2+3m-6}{l}$, with varying
values of $m$ as follows: solid, $m=2$; dashed, $m=3$; dotted,
$m=4$.}
\end{figure}
This shows that energy density of the anisotropic DE is reduced by
the magnetic field. This turns out to be infinite at the initial
epoch, its value decreases with the increase in time and converges
to zero as $T\rightarrow{\infty}$ with the condition that
$n<\frac{3m^2+3m-6}{l}$ shown in Figure \textbf{1(a)}. However, for
$n>\frac{3m^2+3m-6}{l}$, energy density decreases after big bang but
it starts increasing and becomes infinite as $T\rightarrow{\infty}$
shown in Figure \textbf{1(b)}.

Using Eqs.(\ref{26}) and (\ref{31}) in Eqs.(\ref{14}) and
(\ref{15}), the deviation parameters $\delta(T)$ and $\gamma(T)$ become
\begin{eqnarray}\label{32}
\delta(T)=\frac{2n(m+2)(cT^{\frac{-2(3m^2+3m-6-nl)}{3(m-1)}}-
\frac{6{\beta}T^{-4}}{3m^2-3m-nl})}{3(c(2m+1)T^\frac{-2(3m^2+3m-6-nl)}{3(m-1)}-
\frac{{\beta}T^{-4}(3m^2+9m+6-nl)}{3m^2-3m-nl})},\\\label{33}
\gamma(T)=-\frac{nm(m+2)(cT^{\frac{-2(3m^2+3m-6-nl)}{3(m-1)}}-
\frac{6{\beta}T^{-4}}{3m^2-3m-nl})}{3(c(2m+1)T^\frac{-2(3m^2+3m-6-nl)}{3(m-1)}-
\frac{{\beta}T^{-4}(3m^2+9m+6-nl)}{3m^2-3m-nl})}.
\end{eqnarray}
The deviation free EoS parameter can be obtained by using the
expressions for directional Hubble parameters and energy density
in Eq.(\ref{13})
\begin{equation}\label{34}
\omega(T)=-1-\frac{(\frac{4{\beta}T^{-4}(3m^2+9m+6-nl)}{3m^2-3m-nl}-
\frac{2c(1+2m)(3m^2+3m-6-nl)T^\frac{-2(3m^2+3m-6-nl)}{3(m-1)}}{3(m-1)})}
{(2+m)(c(1+2m)T^\frac{-2(3m^2+3m-6-nl)}{3(m-1)}-\frac{{\beta}T^{-4}(3m^2+9m+6-nl)}{3m^2-3m-nl})}.
\end{equation}
The anisotropy measure of anisotropic fluid is given by
\begin{equation}\label{34a}
\frac{(\delta-\gamma)}{\omega}=\frac{n(m+2)^3(cT^{\frac{-2(3m^2+3m-6-nl)}{3(m-1)}}
-\frac{6{\beta}T^{-4}}{(3m^2-3m-nl)})}
{\frac{c(1+2m)(3m^2+3m-6-2nl)}{(m-1)}T^{\frac{-2(3m^2+3m-6-nl)}{3(m-1)}}
+\frac{3\beta(m-2)(3m^2+9m+6-nl)T^{-4}}{{3m^2-3m-nl}}}.
\end{equation}
\begin{figure}
\centering \epsfig{file=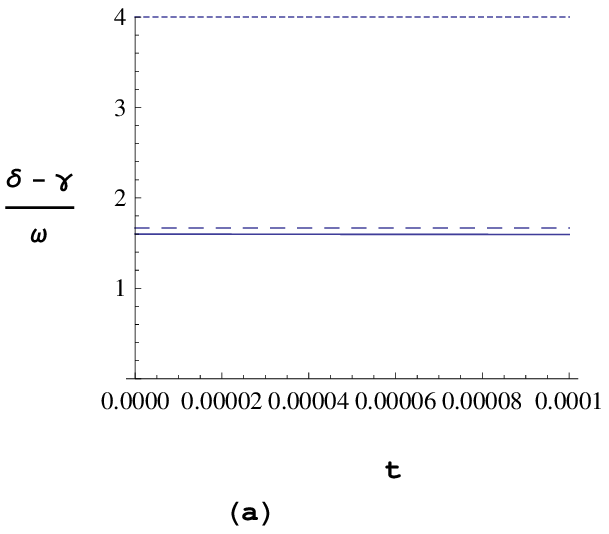,width=.48\linewidth, height=3in}
\epsfig{file=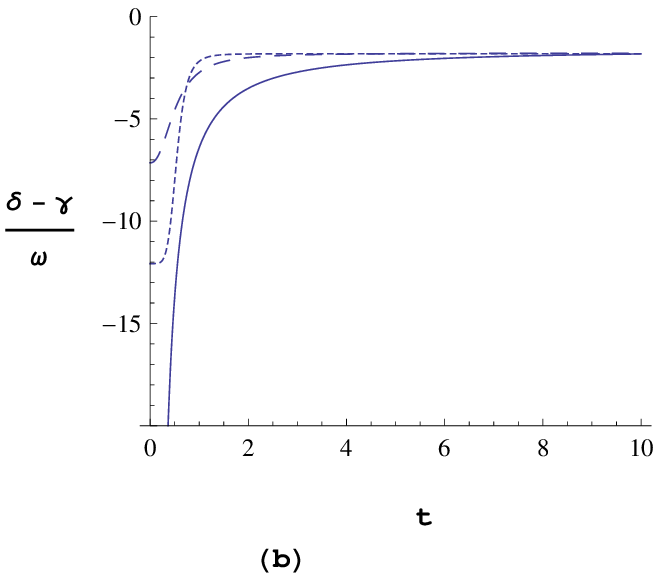,width=.48\linewidth, height=3in} \caption{Plots
of $(\delta-\gamma)/\omega$ verses cosmic time $t$ with
$n<\frac{3m^2+3m-6}{l}$ (a) for initial epoch (b) for future
evolution, with varying values of $m$ as follows: solid, $m=2$;
dashed, $m=3$; dotted, $m=4$.}
\end{figure}

The deviation parameters $\delta(T)$ and $\gamma(T)$ are found to be
finite at $t=0$ and converges to $4n(m+2)/(3m^2+9m+6-nl)$ and
$-2nm(m+2)/(3m^2+9m+6-nl)$ respectively as $T\rightarrow{\infty}$
and $n<\frac{3m^2-3m}{l}$. The anisotropy measure of the DE
$(\delta-\gamma)/\omega$ is constant at $t=0$ shown in Figure
\textbf{2(a)} and converges to $2n(2+m)^3/((2-m)(3m^2+9m+6-nl))$ for
the later times of the universe (Figure \textbf{2(b)}) with the
condition that $n<\frac{3m^2-3m}{l}$ and vice versa for
$n>\frac{3m^2-3m}{l}$. We note that the anisotropy of DE does not
vanish throughout the evolution of the universe. Now we check the
behavior of $\omega$ for $n<\frac{3m^2-3m}{l}$. For the earlier
times of the universe, the deviation free EoS parameter of the DE is
given by $\omega=-1+\frac{2(3m^2+3m-6-nl)}{3(m^2+m-2)}$ which shows
that expansion in the universe may be in the quintessence region
after big bang. Also, ${\omega}\rightarrow{-1+\frac{4}{m+2}}$ as
$T\rightarrow{\infty}$. Hence for later times, $\omega$ represent
EoS of cosmological constant for $m=2$ and $\omega$ may result in
quintessence region for $m>2$. Thus the model represents the
accelerating expanding universe.

\section{Some Special Cases}

Here we discuss the following two special cases, i.e., model with
$\beta=0$ and model with $m=1$.

\subsection{Model with $\beta=0$}

In the absence of magnetic field i.e., $\beta{\rightarrow}0$, the model
(\ref{24}) reduces to the following form
\begin{equation}\label{35}
ds^{2}=\frac{dT^2}{cT^{-\frac{2(3m^2-3-nl)}{3(m-1)}}}
-T^{2m}dx^2-T^2(dy^2+dz^2).
\end{equation}
For this model, the directional, mean Hubble parameters and energy
density of the DE reduce to the following form
\begin{eqnarray}\label{36}
H_x&=&mH_y=c^{1/2}mT^\frac{-(3m^2+3m-6-nl)}{3(m-1)},\nonumber\\
H&=&\frac{c^{1/2}(m+2)}{3}T^\frac{-(3m^2+3m-6-nl)}{3(m-1)},\\\label{37}
\rho&=&c(2m+1)T^\frac{-2(3m^2+3m-6-nl)}{3(m-1)}.
\end{eqnarray}
We see that dynamical $H,~H_x,~H_y$ and $\rho$ are infinite
for earlier times and converge to zero as $T\rightarrow{\infty}$
provided that $n<\frac{3m^2+3m-6}{l}$ and vice versa. The
expansion and scalar are found to be
\begin{eqnarray}\label{38}
\Theta&=&3H=\frac{c^{1/2}(m+2)}{3}T^\frac{-(3m^2+3m-6-nl)}{3(m-1)},\\\label{39}
\sigma&=&\frac{c(m-1)^2}{3}T^\frac{-2(3m^2+3m-6-nl)}
{3(m-1)}.
\end{eqnarray}
The expansion in the universe is infinite at the initial epoch and
decreases with the increase in time for $n<\frac{3m^2+3m-6}{l}$.
However, if $n>\frac{3m^2+3m-6}{l}$ then the universe starts
expanding at $T=0$ and expands indefinitely as
$T\rightarrow{\infty}$. The anisotropy parameter of expansion and
anisotropy measure of fluid become
\begin{eqnarray}\label{39a}
\Delta&=&2\frac{(m-1)^2}{(m+2)^2},\\\label{39b}
\frac{(\delta-\gamma)}{\omega}&=&\frac{c(m-1)(m+2)^3}{(2m+1)(3m^2+3m-6-2ln)}.
\end{eqnarray}
We see that the anisotropy of the expansion and that of DE are
constant so that both model and DE fluid remain anisotropic. The
deviation free EoS parameter is found to be ${1-(2ln)/(3(-2+m+m^2))}$
which is a constant. $\omega$ may begin in the
quintessence or phantom region depending on the values of
constants.

\subsection{Model with $m=1$}

For this value of $m$, Eq.(\ref{20}) implies that
\begin{equation}\label{40}
A(t)=B(t)=a(t)
\end{equation}
and the spacetime (\ref{1}) becomes
\begin{equation}\label{41}
ds^{2}=dt^2-a^2(t)(dx^2+dy^2+dz^2)
\end{equation}
which is the spatially flat FRW metric representing homogeneous and
isotropic universe. Using Eq.(\ref{40}) in (\ref{21}), we have
\begin{equation}\label{42}
3n\frac{\dot{a}^2}{a^2}-\frac{2\beta}{a^4}=0
\end{equation}
which yields
\begin{equation}\label{43}
a(t)=\sqrt{2}({\sqrt{\frac{2\beta}{3n}}t+c})^{1/2}.
\end{equation}
The Hubble parameter and expansion scalar for this model become
\begin{eqnarray}\label{44}
H=\frac{\dot{a}}{a}=\frac{\beta}{(2{\beta}t +c\sqrt{6n\beta})},
\\\label{45} \Theta=3H=\frac{3\beta}{(2{\beta}t
+c\sqrt{6n\beta})}
\end{eqnarray}
while the energy density is
\begin{equation}\label{46}
\rho=\frac{3\beta(2-n)}{8{\beta}t^2+12nc^2+8ct\sqrt{6n\beta}}.
\end{equation}

We find that Hubble parameter and expansion scalar are constant at
initial epoch i.e, at $t=0$ and decrease with the increase in time.
Hence the universe is expanding for the earlier times of the
universe. The energy density is also constant for the earlier times
of the universe and decreases with the increase in time. The
deviation free EoS parameter is $1/3$ which represents the radiation
dominated phase of the early universe \cite{23,34}. Equation
(\ref{43}) implies that for the radiation dominated universe
$a(t){\propto}t^{1/2}$ \cite{35,36}, the model represents expanding
universe. The anisotropy parameter of the expansion is zero since
Eq.(\ref{41}) is an isotropic model. The deviation parameters
$\delta(t),~\gamma(t)$ are constants and hence the anisotropy of the
DE $\frac{\delta-{\gamma}}{\omega}=\frac{6n}{(2-n)}$ is also
constant. The anisotropy of the DE vanishes if we choose the
dimensionless constant $n$ to be zero.

\section{Summary and Conclusion}

In this paper we have constructed Bianchi $I$ cosmological model
with magnetized anisotropic DE fluid having anisotropic EoS. The
deviation parameters $\delta$ and $\gamma$ are obtained by assuming
that conservation equation of DE consists of two separate conserved
parts. The exact solution of the field equations is obtained using
the condition that expansion scalar $\Theta$ is proportional to
$\sigma$. We have discussed some physical aspects of the model both
in the presence and absence of magnetic filed.

The component of magnetic field reduces the energy density,
expansion, shear scalar and Hubble parameters. The expansion in the
universe is found to be infinite at the initial epoch which
decreases with the increase in time. In the absence of magnetic
field, a similar behavior of expansion is observed for
$n<\frac{3m^2+3m-6}{l}$, however, if $n>\frac{3m^2+3m-6}{l}$ then
the universe starts expanding at $T=0$ and expands indefinitely as
$T\rightarrow{\infty}$. The anisotropy measure of the DE is
dynamical and found to be finite for both earlier and later times of
the universe. The isotropic DE can be recovered by choosing $n$ to
be null, where $n$ parameterizes the deviation parameters. The
universe model does not approach to isotropy since anisotropy
parameter of expansion $\Delta$ is found to be constant with and
without magnetic field. For $m=1$, the Bianchi $I$ model is reduced
to spatially flat FRW metric which represents homogeneous and
isotropic universe for earlier times. For this case, the deviation
free EoS parameter is found to be $1/3$ which represents the
radiation dominated phase of the early universe \cite{23,34}.

\end{document}